\documentstyle[aps,prl,epsf]{revtex}

\begin{document}

\draft
\twocolumn[\hsize\textwidth\columnwidth\hsize\csname
@twocolumnfalse\endcsname
\title{Flux flow noise and braided rivers of superconducting vortices}
\author{Kevin E. Bassler$^{1}$, J\"orn Davidsen$^{2,3}$, and Maya 
Paczuski$^{2}$}
\address{$^{1}$Department of Physics, 617 Science \& Research Building I,
University of Houston, Houston, Texas, 77204-5005\\
$^{2}$Department of Mathematics, Imperial College of Science, Technology,
and 
Medicine, London, UK SW7 2BZ\\
$^{3}$Institut f\"ur Theoretische Physik und Astrophysik,
Christian-Albrechts-Universit\"at,
Olshausenstra\ss e 40, 24118 Kiel, Germany}
\date{\today}
\maketitle

\begin{abstract}
Current-voltage measurements of type-II superconductors are described
by a coarse-grained model of  superconducting vortex dynamics.  We
find that the power spectra of the voltage fluctuations, and the noise
power, are related to the large scale morphology of the plastic flux
flow. At currents corresponding to the peak in differential
resistance,  the flux flow forms a braided river, the noise power
is maximal, and the power spectra has a $1/f^{\alpha}$ form with
$\alpha \approx 1.8$ over a wide frequency range.  This agrees with
recent experiments on $NbSe_2$. The observed variation of $\alpha$
with applied current is  a crossover phenomenon.
\end{abstract}
]

\narrowtext

Temporally correlated fluctuations with a power spectrum characterized
by a frequency distribution of the form $1/f^{\alpha}$ appear
ubiquitously in nature.  These correlations are often observed in time
series of nonequilibrium systems such as fluvial rivers
\cite{mandelbrot69}, nerve cells \cite{teich89}, quasars
\cite{press78}, and Barkhausen emission \cite{tadic99}, to name a few.
Despite attempts at a general explanation \cite{btw,soc}, the concrete
sources of these fractal temporal fluctuations are usually unknown.
Here we show, using a cellular model of vortex dynamics, that the
$1/f^\alpha$ spectrum observed in voltage noise measurements of type
II superconductors is related to the complex spatio-temporal pattern
of the plasticly flowing magnetic flux through the system.  This
suggests a general link between long range spatial correlations
observed in driven dissipative systems and long range temporal
correlations obtained in time series measurements of them.  This link
was first proposed in the context of sandpile models \cite{btw}, but
the original models did not display nontrivial $1/f^{\alpha}$ noise
\cite{soc}.  The sandpile type model
presented here to describe plastic flux flow, however, clearly
demonstrates this link.

Marley, Higgins, and Bhattacharya (MHB) found in current-voltage $(IV)$
measurements on the type-II superconductor $NbSe_2$ that the voltage
fluctuations show a $1/f^{\alpha}$ decay in the power spectrum with an
exponent $\alpha$ that varied with the applied transport
current\cite{marley95}. Holding the applied magnetic field constant
and increasing the transport current, they found that $\alpha$ varied,
increasing from its initial value near 1 to about 1.8, and then
falling again.  The low frequency noise power $fS(f)$ was also
measured.  At a constant applied field, $fS(f)$ has a pronounced
maximum at the transport current for which the largest value of
$\alpha$ was measured, suggesting that both phenomena are related.

Those experiments were performed in the so-called
``peak-effect regime'' where
the soft flux line lattice formed by the superconducting vortices is
thought to deform by tearing, giving plastic flow \cite{marley95,higgins96}.  
At magnetic fields $H_{pl}<H<H_p$, the critical current,
$I_c$, is an increasing function of magnetic field, and the $IV$
relation is highly nonlinear.  As $H$ is increased, the size of the
peak in the differential resistance, $dI/dV$, grows, reaching a
maximum value at $H_c \sim 1.88 \; T$ which is significantly different
from $H_{pl} \sim 1.81 \; T$ and $H_p \sim 1.95 \; T$.  The magnitude
of the differential resistance peak shrinks and it occurs at
increasingly larger $I$, with increasing $H>H_c$.  At $H_c$ itself,
the peak in the differential resistance occurs at $I_c$, which marks
the onset of the flow of superconducting vortices where the voltage,
$V$, becomes non-zero.

In this letter, we describe the $IV$ noise experiments of MHB using a
coarse-grained cellular model of superconducting vortex dynamics
\cite{bassler98}, reminiscent of a sandpile model.  We show that a
fractal power spectrum exists only at the maximum of the differential
resistance while for other values of the applied transport current,
$I$, significant deviations from the power-law behavior occur. This
can be easily understood in terms of the underlying morphology of the
vortex flow, which has been shown to form a braided river at the
current corresponding to the peak in the differential resistance
\cite{bassler01}.  The braided structure is also responsible for the
maximum of the low frequency noise power.  In particular,
our results support claims \cite{merithew96}
that complex patterns of the vortex flow
(which were not experimentally observed themselves) are responsible
for the observed noise in the $IV$ measurements.

The cellular vortex model \cite{bassler98} used here is defined as follows.  
Consider a square lattice of size $L \times L$ in which each site $i$ is
occupied by an integer 
number of vortices,
$m_i$.  The total energy of the vortex system includes the
short-ranged repulsive pairwise interaction between the vortices
and the attractive interaction of vortices
with the pinning potential, ${\hat V}$.
For a given configuration of vortex number
$\{m_i\}$, the total energy of the system is:
\begin{equation}
H\bigl(\{m_i\}\bigr)= \sum_{i,j}J_{ij}m_im_j - \sum_i
{\hat V}_i m_i \, \,  ,
\label{hamiltonian}
\end{equation}
where $J_{ij}$ includes
an on-site interaction, and a weaker nearest-neighbor interaction.
The model describes a vortex system
coarse-grained to the scale of the London
length.

The motion of the vortices is assumed to be highly overdamped, described
by the equation of motion $\vec{v} \propto \vec{f}$. The force to move a
unit vortex from $x$ to $y$ is calculated by taking a discrete gradient of
Eqn.~(\ref{hamiltonian}),
\begin{eqnarray}
F_{x \rightarrow y}
& = & {\hat V}_y - {\hat V}_x + [m_x - m_y -1] \nonumber \\
& & + r[m_{x1} + m_{x2} + m_{x3} - m_{y1} - m_{y2} - m_{y3}] \, ,
\end{eqnarray}
where the nearest neighbors of $x$ are $y$,
$x1$, $x2$, and $x3$, and the nearest neighbors cells of $y$ are $x$,
$y1$,
$y2$, and $y3$.
The onsite vortex interactions have unit strength, while $r$ describes the
strength of the nearest neighbor vortex interactions. The normalized
pinning potential ${\hat V}_x$
is a random number taken from a
uniform distribution in the interval between zero
and $V_{max}$. 
In each time step, all cells of the lattice are updated in parallel. A
single vortex moves from a cell to a neighboring cell if the force in
that direction is positive, or equivalently if the total energy of the
system is lowered.

Many alternatives exist to handle the situation when more than one
unstable direction appears for a vortex to move.  In this work, we use
two of them. In the first version of the model, the unstable direction
that has the largest force is chosen and the vortex moves in that
direction. The fluctuation spectrum in this case is generated solely
by a deterministic dynamics.  The second version of the model is
stochastic.  If more than one unstable direction exists, one of them
is chosen randomly and the vortex moves in that direction.

Since the parameter $r$ controls the strength of the vortex
repulsions, increasing $r$ decreases the steady state slope of the
vortex pile. Thus, $r$ controls the critical current, which in the
simulations is the slope of the vortex pile where vortices begin to
continuously move. In the experiments, the critical current
can be controlled by the applied magnetic field.  Therefore,
increasing the applied field $H_{pl}<H<H_p$ in the experiments is
represented by decreasing $r$ in the simulations.  
Periodic boundary conditions apply at the top and bottom
boundaries.

Initially, the system was prepared with a V-shaped flux density
profile characteristic of the Bean state. This describes the
application of a magnetic field, and was accomplished by starting with
an empty lattice and then raising the left and right boundaries to
equal heights, letting vortices enter and penetrate into the system
until a stable state was reached. Then, a transport current $I$ was
applied. According to Ampere's law, a transport current will induce a
magnetic field that raises the magnetic field on one boundary and
lowers it on the opposite side. Thus, a transport current is modeled
by shifting the height of, say, the left boundary verses the right
boundary.  Those changes were made in half integer steps. At small
$I$, the shift between the left and right boundary conditions is
small, and the system reaches a pinned state with no vortex
flow. However, at larger $I$, the average slope is steep enough that
vortices continue to flow from the high (left) boundary to the low
(right) boundary, where they are removed from the system.  The $IV$
characteristic then is the relation between the magnitude of the shift
(representing the applied transport current) and the average flow rate of
vortices (representing the voltage) when the critical current is
exceeded.

\begin{figure}
\narrowtext
\epsfxsize=3.0truein
\centerline{ \epsffile{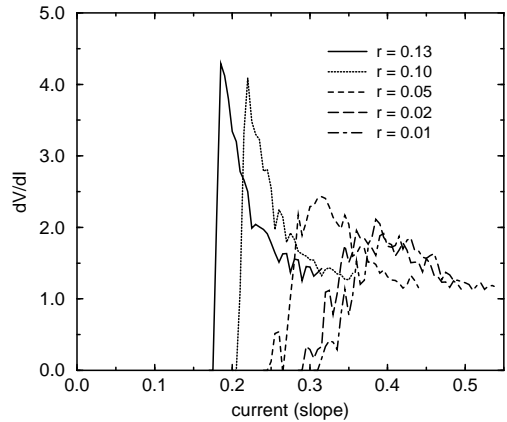} }
\caption{
Simulation results for 
the differential resistance, for five different value
of the vortex interaction strength, $r$. 
}
\end{figure}  

The  differential resistance ($dI/dV$) curves for a number of
different values of $r$ are shown in Fig.~1 \cite{technical} for the
deterministic model. For
$r=0.13$, the $IV$  curve is entirely convex, so the corresponding
differential resistance curve has a large peak at the onset of vortex
motion. This value of $r$ thus corresponds to $H_c$ in the experiments of
MHB. As $r$ decreases, the $IV$ curve becomes S-shaped, and the peak in the
differential resistance gets smaller and occurs at increasing
$I$. These results are entirely consistent with the experimental $IV$
measurements in the range $H_c < H < H_p$.  As shown in
\cite{bassler01}, the shape of the $IV$ curve is related to the
large scale morphology of the vortex flow.  In particular, the peak of the
differential resistance, or equivalently the inflection point of the
$IV$ curve, occurs where the vortex flow has a braided river morphology
which is characterized by a self-affine (multi)fractal structure.
At small $r$ the curve is quite noisy,
especially near onset. This phenomena has been seen in experiments
where it has been called ``IV fingerprints''. As will be discussed
below, this noisiness is due to glassy channel switching of the vortex
flow.

The power spectra for a few values of the applied current at $r=0.13$
are shown in Fig.~2a.  The current $I=0.185$ is just above onset,
where for this value of $r$ the vortex flow is a braided river. The
power spectrum for this value of $I$ shows $1/f^{\alpha}$ type
behavior with $\alpha \approx 1.8$ over a range of frequencies from
about $5.0 \times 10^{-4}$ to about $5.0 \times 10^{-2}$. At higher
frequencies, there is a crossover to behavior that decays less rapidly
with $f$, presumably due to effects associated with the discrete
lattice used in the simulations.  At lower frequencies, there is a
crossover to white noise which is related to the longest time scale
over which correlations are retained in the system.

\begin{figure}
\narrowtext
\epsfxsize=5.1truein
\centerline{ \epsffile{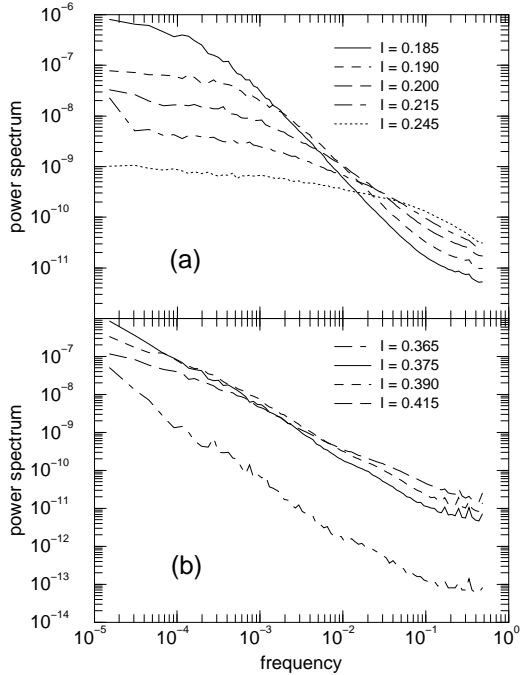} }
\caption{
Voltage noise power spectra for (a) $r=0.13$ and (b) $r=0.01$.
In both graphs, the solid line corresponds to the current
at the peak of the differential, and the straight portion
of both of those curves has a slope of approximately -1.8.
}
\end{figure}  

As $I$ increases, the range of frequency where $1/f^{\alpha}$ noise
occurs shrinks and the white noise roll off at low frequencies
 sets in at a higher
frequency.  This indicates that the long range correlations in the
system fade away. That happens presumably because, as the braided
river floods, which occurs as $I$ increases, the spatial correlation
length shrinks.  The frequency at which the high frequency crossover
occurs is independent of $I$, but the magnitude of the power spectrum
at high frequency grows with $I$ because of the larger number of
vortices moving. If one were to fit a value for the exponent $\alpha$
as a function of $I$, a varying exponent would be measured due to
crossover. The exponent $\alpha$ would range from about 1.8 near the
peak in the differential resistance to near 0 at large $I$. This range
of $\alpha$ is precisely what was reported by MHB close to $H_c$.
Agreement is also found for the low frequency noise power which is
obviously largest for $I=0.185$, and decreases with increasing $I$.
In particular, the crossover picture is consistent with the very
limited range ($\approx 1$ decade) of power-law behavior found in
experiments \cite{rabin98}.  Note that for $r=0.13$, both the
deterministic and stochastic versions of the model give essentially
the same results.

Mostly similar results are also found for the power spectra at $r=0.01$,
which, for the deterministic version of the model, are shown in
Fig.~2b. However, there are also some differences. At this value of
$r$, the peak in the differential resistance does not occur at onset,
which is at $I_c=0.310$, but rather at $I=0.375$. Near $I=0.375$, the
power spectrum has a $1/f^{\alpha}$ form with $\alpha \approx 1.8$
over a broad range of frequencies. There are again crossovers at both
large and small frequencies to regions where the power spectrum decays
less rapidly with $f$.  The frequency at which the low frequency
crossover occurs gets larger with increasing $I$, as the braided river
floods and the spatial correlation length decreases.

\begin{figure}
\narrowtext
\epsfxsize=3.0truein
\centerline{ \epsffile{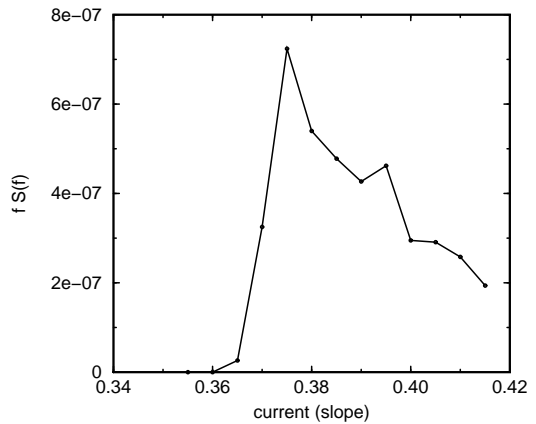} }
\caption{
Low frequency noise power for $r=0.01$. This can be compared with
measurements described in Fig. 4 of Ref. [7].
}
\end{figure}

The low frequency noise power as a function of $I$ for $r=0.01$ is
shown in Fig.~3.  It is calculated by adding all of the points of
$S(f)$, with $f$ between $10^{-5}$ and $10^{-4}$.  The
result rises many orders of magnitude and peaks where the river flow
is braided, at the peak in the differential resistance. This result is
also seen in the experiments.

At transport currents $I$ closer to $I_c$, where the S-shaped IV curve
is concave, and the vortices flow through isolated filamentary
channels \cite{bassler01}, the voltage fluctuations are periodic in
the deterministic model. Also, in this regime, long-time scale
transient and glassy behavior occurs. This is due to channel
switching. At $I$ close to $I_c$, the system typically switches
between a number of different sets of channels before settling into
the steady-state.  The system can continue switching for $10^6$
timesteps, or more, before settling. It does finally settle into a
stationary configuration, but only after many different sets of
channels are probed.  As the current increases, the transient time
becomes longer than we can measure, possibly infinite, near the peak in
the differential resistance.

For the stochastic model, the results at $r=0.01$ differ somewhat from
those of the deterministic model. First, the critical current $I_c$ is
larger in the stochastic case. Apparently, the stochastic lattice
updating can frustrate flow through isolated filamentary
channels. Second, in the stochastic model for $I> I_c$ the system
never settles into a stationary river configuration. Instead, channel
switching continues to occur for at least $10^7$ lattice updates, and
presumably continues indefinitely. However, $1/f^{\alpha}$ noise with
$\alpha \approx 1.8$ also occurs in this version of the model near the
peak in the differential resistance.

Additionally, it is interesting to note that, in the braided regime,
the change in the total number of vortices contained within the system
also show $1/f^\alpha$ fluctuations with $\alpha \approx 1.15$. This
agrees quantitatively with the fall-off power spectrum
measured in the field ramping experiments of Field and co-workers
\cite{field95}.

Other theoretical studies of vortex dynamics have also found
$1/f^{\alpha}$ behavior in the power spectrum of the voltage noise.
Driving the system in a different fashion designed to maintain a
critical state, Mohler and Stroud \cite{mohler99} using the cellular
model \cite{bassler98} and Olson, et al. \cite{olson98}, using
molecular dynamics simulations found fractal behavior in the noise
spectrum. Those results are consistent with the fact that in both
cases the vortex flow has a braided river morphology
\cite{olson98,bassler99}.  However, the value of $\alpha$ they
measured deviates from that measured here
due to the different driving. 
Another approach was taken by Dom\'{\i}nguez \cite{dominguez99}.
Integrating a set of equations that results from a
discretization of the London model, he
found two regimes of vortex flow, filamentary motion near onset, and
homogeneous motion at higher currents. In the filamentary channel
regime, he discovered $1/f$ behavior of the voltage noise power
spectrum. 
Using molecular dynamics simulations, Kolton et
al. \cite{kolton99}, also found the same two regimes and a peak in the
low frequency noise power at the transition between them. They also
suggested that the transition is from plastic flow in the filamentary
regime to smectic flow in the homogeneous regime, thereby associating
the transition with a reordering of the flux line lattice. The cellular
model used here cannot describe any possible
structural transitions in the flux line lattice. Nevertheless, we find
remarkable quantitative similarity with experimental voltage noise
measurements involving the transition from filamentary, through
braided, to homogeneous motion of the vortices.

In summary, voltage noise in $IV$ measurements has been studied in a
simple, cellular model of vortex dynamics. The model captures the
essential features of the plastic flow of the flux line lattice formed
by the superconducting vortices: $1/f^{\alpha}$ behavior of the power
spectrum was observed where the differential resistance has a peak,
and the vortex flow has been shown to form a braided river. This
result is robust, occurring over all values of the vortex interaction
strength $r$, and for two different versions of the model, one
completely deterministic and the other stochastic. The results model
experiments conducted on $NbSe_2$, in the peak-regime where the FLL
moves plasticly.  Most notably, quantitative agreement is found in the
behavior of the exponent $\alpha$ between simulations and experiments,
both of which find $\alpha \approx 1.8$ at the peak of the
differential resistance. Both also find that the low frequency noise
power is maximal at the peak in the differential resistance.
Therefore, we conclude that is the self-organized spatio-temporal structure
of the braided river flow that is responsible for the observed
$1/f^{\alpha}$ behavior of the $IV$ power spectra in superconductors.

This work was supported by the NSF through grant \#DMR-0074613. KEB also thanks
the Alfred P. Sloan Foundation and the Texas Center for Superconductivity
for support. JD would like to thank the DAAD for
financial support and the Imperial College for support and hospitality.


\begin{references}


\bibitem{mandelbrot69}
B.~B. Mandelbrot and J.~R. Wallis,  Water Resour. Res. {\bf 5},  321
(1969).

\bibitem{teich89}
F. Gr{\"u}neis {\it et~al.}, Biol. Cybern. {\bf 60},  161  (1989).

\bibitem{press78}
W.~H. Press, Comments Astrophys. {\bf 7},  103  (1978).

\bibitem{tadic99}
B. Tadi{\'c}, Physica A {\bf 270},  125 (1999).

\bibitem{btw} P. Bak, C. Tang, and K. Wiesenfeld, Phys. Rev. Lett. {\bf 59},
381 (1987).


\bibitem{soc}
For a review see P. Bak,
 {\it How Nature Works: The Science of
Self-Organized Criticality} (Copernicus, New York, 1996).


\bibitem{marley95}
A. C. Marley, M. J. Higgins, and S. Bhattacharya, Phys. Rev. Lett. {\bf
74}, 3029 (1995).

\bibitem{higgins96}
M. J. Higgins and S. Bhattacharya, Physica C {\bf 257}, 232 (1996).

\bibitem{bassler98}  
K. E. Bassler and M. Paczuski, Phys. Rev. Lett. {\bf 81}, 3761 (1998).

\bibitem{bassler01} 
 K. E. Bassler, M. Paczuski, and E. Altshuler, to appear in Phy. Rev. B (2001).

\bibitem{merithew96}  
R. D. Merithew {\it et~al.}, Phys. Rev. Lett. {\bf 77} 3197 (1996).

\bibitem{technical}
After the boundaries are shifted, the system with $L = 200$ and
$V_{max}=5.0$ was allowed to run for
$4.5 \times 10^6$ lattice updates before the boundaries were shifted
again. If during that time the system reached a stable state, the voltage
was taken to be zero, otherwise the voltage $V$ was taken to be equal to
the
average number of vortex moves per site in the direction of the right
boundary.
To avoid transient effects the first $10^6$ updates were discarded. 

\bibitem{rabin98}
M. W. Rabin, {\it et al.}, Phys. Rev. B {\bf 57}, 720 (1998).

\bibitem{field95}
S. Field, {\it et al.}, Phys. Rev. Lett. {\bf 74}, 1206 (1995).

\bibitem{mohler99} G. Mohler and D. Stroud, 
Phys. Rev. B {\bf 60}, 9738 (1999).

\bibitem{olson98} C. J. Olson, C. Reichhardt, and F. Nori, 
Phys. Rev. Lett. {\bf 80}, 2197 (1998).

\bibitem{bassler99}  
K. E. Bassler, M. Paczuski, and G. F. Reiter, Phys. Rev. Lett. {\bf 83}, 3965 (1999).

\bibitem{dominguez99} D. Dom\'{\i}nguez, Phys. Rev. Lett. {\bf 82}, 181 (1999).

\bibitem{kolton99} A. B. Kolton, D. Dom\'{\i}nguez, and N. Gr{\o}nbech-Jensen,
Phys. Rev. Lett. {\bf 83}, 3061 (1999).


\end{references}
\end{document}